\documentclass[12pt]{iopart}

\usepackage{amsfonts}

\begin{document}

\jl{6}

\title{The QCD  $M$embrane}

\author{Stefano Ansoldi\dag{}\footnote[3]{e-mail
        address: \texttt{ansoldi@trieste.infn.it}},
        Carlos Castro\ddag{}\footnote[4]{e-mail
        address: \texttt{castro@ctsps.cau.edu}},
        Euro Spallucci\dag{}\footnote[5]{e-mail
        address: \texttt{spallucci@trieste.infn.it}}}

\address{\dag{}Dipartimento di Fisica Teorica
Universit\`a di Trieste, and INFN, Sezione di Trieste}

\address{\ddag{}Center for Theoretical Studies of Physical Systems
Clark Atlanta University, Atlanta, GA.3t314}

\begin{abstract}
   In this paper we study \textit{spatially} quenched, $SU(N)$ Yang--Mills
   theory in the large-$N$ limit. The resulting reduced action shows the
   same formal look as the  Banks--Fischler--Shenker--Susskind
   $M$--theory action. The  Weyl--Wigner--Moyal \textit{symbol}
   of this matrix model is the Moyal deformation  of a $p(\, =2\,)$--brane 
   action. Thus, the large-$N$ limit of the  spatially quenched $SU(N)$ 
   Yang--Mills is seen to describe a dynamical membrane. By assuming spherical 
   symmetry we compute the mass spectrum of this object in the WKB 
   approximation.
\end{abstract}

\pacno{11.17}

\submitted

\maketitle

\section{Introduction}
The pivotal role played by gauge field theories in
a unified description of fundamental interactions proposed one of the most
challenging  questions of modern high energy theoretical physics:
if Nature likes so much gauge symmetry why gravity
cannot fit into such an elegant and, would be, universal blueprint?\\
Before the advent of string theory this was a question  without an answer.
After that, it became clear that all field theories, including Yang--Mills
type models, must be seen as low energy, effective approximations of some
more fundamental theory where the dynamical degrees of freedom are carried by
relativistic extended objects. Furthermore, even the low--energy
effective gauge theories require a ``\textit{stringy}\/'' approach in the
\textit{strong coupling} regime, where standard perturbation series breaks
down. Color confinement in QCD is a remarkable example of a phenomenon
where the string tenet meets gauge symmetry. The stringy aspects of
confinement have long been studied, but are not yet fully
understood\footnote{There are many interesting reviews on this problem,
as \cite{migdal}, \cite{polya}, \cite{antonov}, \dots and
we apologize for omitting many other good ones.}.
Several different models have been proposed as \textit{phenomenological}
descriptions of the quark--gluon bound
states, including color flux tubes, three--string
of various shapes,
bag--models \cite{guidry}. To promote some of them
to a deeper status one would like to derive extended objects
as non-perturbative excitations of an underlying gauge theory  \cite{gauge}.
%%%%%%%%%%%%%%%%%%%%%%%%%%%%%%%%%%%%%%%%%%%%%%%%%%%%%%%%%%%%%%%%%%%%%%%%%%%%%%%
%
%%%%%%%%%%%
The first remarkable achievement of this program was to obtain stringy
objects from $SU(N)$ Yang--Mills theory in the large--$N$ limit \cite{bars}.

These results have been
extended to the case of a self--dual membrane in
the $SU(\infty)$ Toda model \cite{toda}.
In a nutshell, the
problem is to establish a formal correspondence between a Yang--Mills
connection,
$A^i{}_j{}_\mu(x)$, and the string  coordinates, $X^\mu(\sigma^0\ ,\sigma^1)$,
i.e.  one has to ``get rid of'' the internal, non-Abelian,
 indices $i$, $j$   and replace the spacetime coordinates $x^\mu$ with
 two continuous coordinates $\left(\, \sigma^0\ ,\sigma^1\,\right)$.
With hindsight, the ``recipe'' to turn a non-Abelian gauge field into a set
of Abelian functions describing the embedding of the string world-sheet
into target spacetime can be summarized as follows:\\
a) transform the original field theory into a sort of ``matrix quantum
mechanics'';\\
b)  use the Wigner-Weyl-Moyal map to build up the {\it symbol} associated to
the above matrix model;\\
c)  take the ``{\it classical limit}'' of the theory obtained in b).\\

Stage a) requires two sub--steps:\\
a1) take the large--$N$ limit, i.e. let the row and column labels
$i$, $j$ to range over arbitrarily large values;
a2) dispose of the spacetime coordinate dependence through the so called
``quenching approximation'', i.e. a technical manipulation
which is formally equivalent to collapse the whole spacetime to a single point.
\\
After stage a) the original gauge theory is transformed into a quantum
mechanical model where the physical degrees of freedom are carried by
large coordinate independent matrices. Then, stage b) associates to each of
such big matrices its corresponding symbol, i.e. a function defined over
an appropriate non-commutative phase space. The resulting theory is a {\it
deformation} of an ordinary field theory, where the ordinary product
between functions is replaced by a non-commutative $\ast$--product.
The deformation parameter, measuring the amount of non-commutativity,
results to be $1/N$, and the classical limit corresponds exactly to the
large--$N$ limit. The final result, obtained at the stage c), is
a string action of the Schild type, which is invariant under area-preserving
reparametrization of the world-sheet.\\
More recently,
we have also shown that \textit{bag\/}-like objects fit the large-$N$ spectrum 
of
Yang--Mills type theories as well, both in four \cite{noi1} and higher 
dimensions.
We started from the Yang--Mills action for an $SU(\,N\,)$ gauge theory
supplemented by a topological term
\begin{equation}
S\equiv -\int d^4x\, \left(\, {N\over 4g^2_{YM}}\,\mathrm{Tr} \bi{F}_{\mu\nu}
\bi{F}^{\mu\nu}+ {\theta\, N\, g^2_{YM}\over 16\pi^2}\, 
\epsilon^{\mu\nu\rho\sigma}
\, \mathrm{Tr}\bi{F}_{\mu\nu} \, \bi{F}_{\rho\sigma}\,\right)\label{ymcs}
\end{equation}
and went trough the steps from a) to c). As a final result we obtained
the following action
\begin{eqnarray}
W= -{\mu^4_0\over 16}\,\int_\Sigma d^4\sigma\,
\left\{\, X^\mu\ , X^\nu\,\right\}_{PB} \left\{\, X_\mu\ , X_\nu\,\right\}_{PB}
+
\nonumber \\
\qquad
-\kappa\,\epsilon_{\mu\nu\nu\rho\sigma} \,\int_{\partial\Sigma} d^3s
 X^\mu\, \left\{\, X^\nu\ , X^\rho\ ,X^\sigma\right\}_{NPB}
 \quad ,
\label{bag}
\end{eqnarray}
where
$$
\left\{\, X^\mu\ , X^\nu\,\right\}_{PB}\equiv \epsilon^{mn}\,\partial_m
\, X^\mu\,\partial_n\, X^\mu
$$
is the Poisson Bracket and
$$
\left\{\, X^\nu\ , X^\rho\ ,X^\sigma\right\}_{NPB}\equiv \epsilon^{ijk}
\partial_i X^\nu \, \partial_j X^\rho\,\partial_k X^\sigma
$$
is the Nambu--Poisson Bracket.
The first term in (\ref{bag}) describes a bulk three--brane, or bag, which in
four dimensions is a pure volume term characterized by a pressure $\mu^4_0$.
All the dynamical degrees of freedom
are carried by the second term in (\ref{bag}), where $\kappa$ is the
membrane tension; this term encodes the dynamics of
the boundary, Chern--Simons membrane, enclosing the bag. Tracing back the bulk
and boundary terms in the original action (\ref{ymcs}) it is possible to
establish the following formal correspondence
\begin{eqnarray}
\hbox{``glue''} \longleftrightarrow \hbox{bulk $3$-brane}
\nonumber\\
\hbox{``instantons''} \longleftrightarrow \hbox{Chern--Simons boundary 
$2$-brane}
.
\nonumber
\end{eqnarray}
This scheme, which  has been generalized to
Yang--Mills theories in higher dimensional spacetime \cite{noi2}, points out
that not only strings but bag-like objects fit the large--$N$ spectrum
of $SU(N)$ gauge theories. However, a non-trivial dynamics for
these spacetime filling objects comes only from boundary effects,
described by Chern--Simons terms in the original gauge action. \\
Against this background, we would like to investigate the existence
of non-topological membrane-like excitations in the large
large--$N$ spectrum of a $SU( N )$  Yang--Mills theory in four dimensions.
Clues suggesting the existence of these objects come from earlier  Abelian
models \cite{ax} and the recent conjectures about $M$--Theory. $M$(atrix) theory
is the, alleged, ultimate non--perturbative formulation of string theory.
More in detail, two models have been constructed as possible non--perturbative
realization of Type IIA \cite{bfss} and Type IIB \cite{ikkt} string theory.
The matrix formulation of Type IIB strings is provided by a large--$N$,
$10$-dimensional super Yang-Mills theory reduced to a single point
\begin{equation}
S_{IKKT}= -{\alpha\over 4}\, \mathrm{Tr} \left[\,\bi{A}_\mu\ ,
\bi{A}_\nu\, \right]^2+\dots\label{ikkt}
.
\end{equation}
The dots refers to the fermionic part of the action which is not relevant
to our discussion. The model
(\ref{ikkt}) has a rich spectrum of extended objects. Our investigations
in \cite{noi1} and \cite{noi2} has been initially triggered by the formal
analogy between (\ref{ikkt}) in $10D$  and quenched  Yang--Mills theory in 
$4D$.\\
Matrix description of Type IIA strings is given in terms of $0$-branes quantum
mechanics
\begin{equation}
S_{BFSS}= {1\over 2g_s}\, \mathrm{Tr}\, \left(\,{ d\bi{X}^i\over dt}\,
{ d\bi{X}_i\over dt}-{1\over 2}\left[\,\bi{X}^i\ ,
\bi{X}^j\, \right]^2 +\dots \,\right)
,
\label{bfss}
\end{equation}
where $i=1 , \dots , 9$. Again, the $0$-branes matrix coordinates can be
seen as Yang--Mills fields reduced to a line. \\
In this paper
we would like to ``reverse the path'' leading Type IIA strings to the
matrix model (\ref{bfss}) and show how a $3D$ version of (\ref{bfss})
can be obtained from the {\it canonical} formulation of an $SU(\,N\,)$
Yang--Mills theory through a modified quenching prescription. Then, we are 
going to extract a {\it non--relativistic,} dynamical $2$--brane from the 
large--$N$ spectrum of the model by following the procedure introduced in 
\cite{noi1} and \cite{noi2}.\\
%%%%%%%%%%%%%%%%%%%%%%%%%%%%%%%%%%%%%%%%%%%%%%%%%%%%%%%%%%%%%%%%%%%%%%%%%%%%%%%
%

The paper is organized as follows: in section \ref{ymatrix} we start from
$SU(N)$ Yang--Mills theory in four dimensions and obtain a corresponding
matrix theory through the quenching approximation; two different type of
quenching are discussed in section \ref{subsikkt} and section \ref{subsbfss};
in section \ref{moyal} we study the large-$N$ limit of the matrix model
introduced in section \ref{subsbfss} and use the Weyl--Wigner--Moyal map
to get the action for a membrane; we conclude the paper by computing the
mass spectrum of this membrane in the WKB approximation.

\section{Yang--Mills Theory as a Matrix model}
\label{ymatrix}
In the introduction we referred to the supposed relation between confinement
and extended excitations of the Yang--Mills field.  Recently, an even deeper
and more fundamental relation between branes and Yang--Mills fields has
been conjectured in the framework of $M$-theory \cite{banks}. Non-perturbative
formulations of string theory require the introduction of  higher dimensional
solitonic objects satisfying Dirichlet boundary conditions \cite{dbrane}.
Dirichlet-branes are formally described by non-commuting matrix coordinates.
Thus, non-perturbative string theory, or $M$-theory, is conveniently
written in terms of matrix dynamical variables. The corresponding low energy,
effective, supersymmetric Yang--Mills theory is derived through an appropriate
compactification procedure of the original matrix model \cite{bfss},
\cite{tg}.\\
In this section we are going to follow a similar path, but in the opposite
direction: we shall start from a $SU(  N  )$ gauge theory in four dimensional
spacetime and build a matrix model. Our final purpose is to show that the
spectrum of such a QCD matrix model contains dynamical membrane type
objects. We thus start from the Yang--Mills action
\begin{equation}
S_{\mathrm{YM}}=\int dt\int_{V_H} d^3x \,
{\mathcal{L}}_{\mathrm{YM}}\left( \bi{F}\ , \bi{A} \right)
\end{equation}
defined in terms of the Lagrangian density
\begin{equation}
{\mathcal{L}}_{\mathrm{YM}}\left( \bi{F}\ , \bi{A} \right)\equiv
{N\over 4 g_0{}^2}   \mathrm{Tr}  \left(  \bi{F}_{\mu\nu}  \right)^2 -{1\over 2
g_0{}^2}   \mathrm{Tr}  \bi{F}^{\mu\nu} \bi{D}_{[  \mu}  \bi{A}_{\nu ]}
\ .
\label{fo}
\end{equation}
The covariant derivative is defined as usual,
\begin{equation}
\bi{D}_{[  \mu}  \bi{A}_{\nu ]}\equiv \partial _{[  \mu}  \bi{A
}_{\nu ]}+i \left[  \bi{A}_\mu, \bi{A}_\nu \right]
\ ,
\end{equation}
but the $SU( N )$ Yang--Mills Lagrangian
${\mathcal{L}}_{\mathrm{YM}}\left( \bi{F}, \bi{A} \right)$ is written
in the first order formulation: thus
$$
\bi{A}_\mu\equiv A_\mu^a\bi{T}^a,\quad \bi{F}_{\mu\nu}\equiv
 F_{\mu\nu}^a\bi{T}^a
$$
are \textit{independent} vector and
tensor fields respectively\footnote{The metric signature is $(- \, + \, + \, +)
$ and
in our notation matrices are denoted by boldface letters.} valued
in the Lie algebra defined by the commutation relations
$$
\left[  \bi{T}^a,\bi{T}^b \right]=
i  f^{abc}  \bi{T}^c.
$$
The integration
volume $V_H$ will be specified later on.
The form  (\ref{fo}) is appropriate for an Hamiltonian
formulation of the action, as it is required by the quenching approximation
that we shall apply in the next section.\\
Starting from the Lagrangian density we can, as usual,
introduce the canonical momentum and Hamiltonian
\begin{eqnarray}
\bi{E}^i\equiv {\partial {\mathcal{L}}_{\mathrm{YM}}\over \partial \partial_t
\bi{A}_i}
=- {1\over 2g_0{}^2}   \bi{F}^{t  m}
\\
H_0\equiv \mathrm{Tr} \left(  \bi{E}^i  \partial_t  \bi{A}_i \right)-{\mathcal{
L}}_0
\end{eqnarray}
and rewrite (\ref{fo}), in terms of phase space variables, as:
\begin{eqnarray}
\fl
{\mathcal{L}}_{\mathrm{YM}}
&{=} -{N\over 2 g_0{}^2}   \mathrm{Tr}  \left(  \bi{F}^{t m}
\right)^2-{N\over 2 g_0{}^2}   \mathrm{Tr}   \bi{F}^{t m}
\left(  \partial_t   \bi{A}_i +  \partial_i \bi{A}_t
-i  \left[  \bi{A}_t, \bi{A}_i \right] \right)
\nonumber\\
\fl & \qquad
+{N\over 4 g_0{}^2}   \mathrm{Tr}  \left(  \bi{F}_{mn}  \right)^2
-{N\over 2 g_0{}^2}   \mathrm{Tr}  \bi{F}^{mn} \bi{D}_{[ m}  \bi{A}_{nu ]}
\nonumber\\
\fl
&{=}
-{g_0{}^2\over 2N}  \mathrm{Tr} \left(  \bi{E}^i \right)^2 +{1\over 2 }
\mathrm{Tr} \left( \bi{E}^i \partial_t   \bi{A}_i \right)
 -{1\over 2 }  \mathrm{Tr} \left( \bi{A}_t  \bi{D}_i  \bi{E}^i \right)
 -{N\over 4 g_0{}^2}  \mathrm{Tr}  \left(   \bi{D}_{[  m} \bi{A}_{n ]} \right)^
2
 .
\end{eqnarray}
Accordingly, the phase space action reads
\begin{eqnarray}
\fl S
&=
\int \!\!dt\int_{V_H} \!\!\!\!\!\!
d^3x \left[  \bi{E}^i  \partial_t  \bi{A}_i -
H_0 \right]\\
\fl
&=
\int \!\!dt\int_{V_H} \!\!\!\!\!\!
d^3x \left[  {g_0{}^2\over 2}  \mathrm{Tr} \left(
\bi{E}^i \right)^2 \!\! -{1\over 2 }    \mathrm{Tr} \left(
\bi{E}^i \partial_t   \bi{A}_i \right)
 +{1\over 2 }  \mathrm{Tr} \left( \bi{A}_t  \bi{D}_i  \bi{E}^i \right)
% \right.
%\nonumber\\
% \left.
 +{1\over 4 g_0{}^2}  \mathrm{Tr}  \left(   \bi{D}_{[  m} \bi{A}_{n ]} \right)^
2
 \right]
 .
 \label{canon}
\end{eqnarray}
Let us remark that
$\bi{A}_t$ enters linearly in the canonical form of the action (\ref{canon})
and plays the role of Lagrange multiplier enforcing the (non-Abelian) Gauss
Law:
\begin{equation}
{\delta S\over \delta\bi{A}_t }=0\quad\Longrightarrow \quad
\bi{D}_i  \bi{E}^i=0
.
\label{gauss}
\end{equation}
Thus, solving the classical field equation for $\bi{A}_t$ is equivalent as
requiring $\bi{E}^i$ to be covariantly divergence free in vacuum. Thus,
inserting the solution of the Gauss constraint (\ref{gauss}),
the action for $\bi{E}^i$ becomes
\begin{equation}
\fl S_{\mathrm{YM}}=\int dt\int_{V_H} d^3x \left[ {g_0{}^2\over 2N}
\mathrm{Tr} \left(  \bi{E}^i \right)^2 -{1\over 2 }    \mathrm{Tr} \left(
\bi{E}^i \partial_t   \bi{A}_i \right)+
{N\over 4 g_0{}^2}  \mathrm{Tr}  \left(   \bi{D}_{[  m} \bi{A}_{n ]} \right)^2
 \right]
 .
\end{equation}
We will now go on with the quenching procedure.

\subsection{Spacetime Quenching $\longrightarrow$ IKKT--type model}
\label{subsikkt}
%%%%%%%%%%%%%%%%%%%%%%%%%%%%%%%%%%%%%%%%%%%%%%%%%%%%%%%%%%%%
In order to provide the reader a self--contained derivation of our model,
let us briefly review how the quenching approximation works in a simple
toy model \cite{miao}. Consider the following two-dimensional model of matrix
non-relativistic quantum mechanics

\begin{equation}
S\equiv \int d^2x\, \mathrm{Tr}\,\left[\, {1\over 2}\, \left(\, \partial_0 
\bi{M}\,
\right)^2 -{1\over 2}\, \left(\, \partial_1 \bi{M}\,
\right)^2 -V\left(\, \bi{M}\,\right)\,\right],
\end{equation}
where $\bi{M}(\,x^0\ , x^1\,)$ is an Hermitian, $2\times2$ matrix and
$V\left(\,\bi{M}\,\right)$ is an appropriate potential term and we
suppose that the system described by $\bi{M}$ is
enclosed in a (one-dimensional) ``spatial box'' of size $a$.
Quenching is an approximation borrowed from the theory of spin
glasses where only ``slow'' momentum modes are kept to compute the spectrum
of the model. Slow modes are described by the eigenvalues of the
linear momentum matrix $\bi{P}$, while the off-diagonal, ``fast'' modes
can be thought as being integrated out.
Thus, $\bi{M}(\, x\,)$ can be written as
\begin{equation}
\bi{M}(\, x^0\ , x^1\,)= \exp\left(\, i\,\bi{P}\,x^1\,\right)
\bi{M}(\, x^0\,)\exp\left(\, -i\,\bi{P}\,x^1\,\right)
\end{equation}
and the spatial derivative $\partial_1 \bi{M}(\,x^0\ , x^1\,)$ becomes the
the commutator of $\bi{P}$ and $\bi{M}$
\begin{equation}
\partial_1 \bi{M}=i\,\left[\,\bi{P}\ , \bi{M}\,\right]
\end{equation}
%To complete the quenching procedure we restrict the spatial integration
%to a finite range, say $a\equiv 2\pi/\Lambda$, acting as an inverse
%ultraviolet cut-off. The quenching procedure provide us with an action
so that the action becomes
\begin{equation}
S\equiv a\,
\int dx^0\, \mathrm{Tr}\,\left[\, {1\over 2}\, \left(\, \partial_0 \bi{M}\,
\right)^2 +{1\over 2}\, \left[\, \bi{P}\ , \bi{M}\, \right]^2
 -V\left(\,\bi{M}\,\right)\,\right],
\end{equation}
in which the the $x^1$ dependence of the original matrix field has been
removed.\\
%%%%%%%%%%%%%%%%%%%%%%%%%%%%%%%%%%%%%%%%%%%%%%%%%%%%%%%%%%%%
Quenching can be applied to a Yang--Mills gauge  theory by taking into account
that in the large-$N$ limit $SU(  N )\to U(  N )$ and the group of spacetime
translations fits into the diagonal part of $U( \infty )$. By neglecting
off-diagonal components, spacetime dependent dynamical variables can be
shifted to the origin by means of a translation operator $\bi{U}(x)$:
since the translation group is Abelian one can choose the matrix $\bi{U}(x)$
to be a plane wave diagonal matrix \cite{kita}
\begin{equation}
\bi{U}_{ab}(x)=\delta_{ab}\exp\left(  i  q^a{}_\mu  x^\mu \right)
\ ,
\end{equation}
where $ q^a {}_\mu $ are the eigenvalues of the four-momentum $\bi{q }_\mu$.
Then
$$
\bi{A}_\mu(x)=\exp\left(-i\bi{q }_\mu  x^\mu \right)  \bi{A}_\mu(0)
\exp\left(  i\bi{q }_\mu  x^\mu \right)\equiv
\bi{U}^\dagger(x)  \bi{A}_\mu^{(0)} \bi{U}(x)
$$
and in view of the equality
$$
\bi{D}_ \mu  \bi{A}_\nu = i \bi{U}^\dagger(x)
 \left[  \bi{q }_\mu+\bi{A}_\mu^{(0)} ,
\bi{A}_\nu \right] \bi{U}(x),
$$
which when antisymmetrized yields
$$
\bi{D}_{[  \mu} \bi{A}_{\nu ]} = i \bi{U}^\dagger(x)
\left[  \bi{q }_\mu +\bi{A}_\mu^{(0)} ,
  \bi{q }_\nu  +  \bi{A}_\nu^{(0)} \right]\bi{U}(x)\nonumber\\
\equiv i   \bi{U}^\dagger(x)  \left[  \bi{A}_\mu^{(\mathrm{q})} ,
\bi{A}_\nu^{(\mathrm{q})} \right] \bi{U}(x)
,
$$
we can see that the translation is compatible with the
covariant differentiation, so that
$$
\bi{F}_{\mu\nu}(x)= \exp\left(-i\bi{q }_\mu x^\mu \right)  \bi{F
}_{\mu\nu}(0)\exp\left(  i\bi{q }_\mu x^\mu  \right)\equiv
\bi{U}^\dagger(x)  \bi{F}_{\mu\nu}^{(0)}  \bi{U}(x)
.
$$
Once the original gauge field theory is turned into a constant matrix model,
we still need to dispose of the spacetime volume integration.
The gluon field is spatially confined inside a volume $V_H$ comparable with the
typical size of an hadron. Thus, for any finite time interval $T$
we can replace the four-volume integral by
$$
\int_0^T  dt\int_{V_H} d^3x\quad\longrightarrow \quad T\, V_H
$$
and the quenched action becomes
\begin{equation}
S^{(\mathrm{q})}_{\mathrm{YM-red.}}= T\,  V_H
{N\over g_0{}^2 }  \mathrm{Tr}
 \left(  {1\over 4}\left(  \bi{F}_{\mu\nu}^{(0)}  \right)^2 -{i\over 2}
 \bi{F}^{(0)  \mu\nu} \left[  \bi{A}_\mu^{(\mathrm{q})} ,
\bi{A}_\nu^{(\mathrm{q})}  \right] \right)
\ ,
\label{eq:qYMred}
\end{equation}
which is the first order formulation of the IKKT--type action in four
spacetime  dimensions \cite{ikkt}.
The usual second order formulation is readily obtained by
solving for $\bi{F}_{\mu\nu}^{(0)}$ in terms of $\bi{A}_\mu^{(\mathrm{q})}$
$$
\bi{F}_{\mu\nu}^{(0)}= i  \left[  \bi{A}_\mu^{(\mathrm{q})} ,
\bi{A}_\nu^{(\mathrm{q})} \right]
$$
and substituting back this result into (\ref{eq:qYMred})
\begin{equation}
S^{(\mathrm{q})}_{\mathrm{YM-red.}}\rightarrow S^{IKKT}_{(4)}= \beta  V_H
   {N\over 4g_0{}^2 }  \mathrm{Tr}
\left(  \left[  \bi{A}_\mu^{(\mathrm{q})} , \bi{A}_\nu^{(\mathrm{q})} \right]
\right)^2\ .
\label{sikkt}
\end{equation}

The string-like excitations  of this model and the relation between large-$N$
gauge symmetry and area--preserving diffeomorphism
have been investigated in several papers
\cite{bars}. More recently, we found that not only  strings
are  present in the large-$N$ spectrum of (\ref{sikkt}) but also
spacetime filling, bag-like objects \cite{noi1}, for which a non-trivial
boundary
dynamics was found
through the addition of topological terms to the original
Yang--Mills action. Here, we would like to explore a different route leading in
a more straightforward way to a dynamical brane action. From this purpose we
need to introduce a different quenching approximation, which we discuss in the
next section.

\subsection{Spatial Quenching $\longrightarrow$ BFSS--type model}
\label{subsbfss}
Instead of shifting $\bi{A}_\mu(  t,\vec x )$ to a single point, as
we did in the final part of the previous section, we translate the matrix
gauge field to a fixed time slice by means of the conserved three--momentum
$\vec q$. As in the previously discussed case
\begin{eqnarray}
\bi{A}_i(t, \vec x)=\exp\left(-i\bi{q }_i  x^i \right)
\bi{A}_i(t)  \exp\left(  i\bi{q }_i  x^i \right)\equiv
\bi{U}^\dagger(\vec x)  \bi{A}_i(t) \bi{U}(\vec x)
\nonumber \\
\bi{A}_t(t, \vec x)=\exp\left(-i\bi{q }_i  x^i \right)
\bi{A}_t(t)  \exp\left(  i\bi{q }_i  x^i \right)\equiv
\bi{U}^\dagger(\vec x)  \bi{A}_t(t) \bi{U}(\vec x)
,
\nonumber
\end{eqnarray}
the translation operation commutes with the covariant differentiation since
$$
\bi{D}_i  \bi{A}_n = i \bi{U}^\dagger(\vec x)
\left[  \bi{q }_i + \bi{A}_i(t) ,
\bi{A}_n \right] \bi{U}(\vec x)
$$
implies
$$
\bi{D}_{[  m} \bi{A}_{n ]} = i \bi{U}^\dagger(\vec x)
\left[  \bi{q }_i +\bi{A}_i(t) ,
\bi{q }_n  +  \bi{A}_\nu(t) \right]\bi{U}(\vec x)
$$
and for the conjugate momentum we also get
$$
\bi{U}^\dagger(\vec x)  \bi{E}_i(t) \bi{U}(\vec x)
\equiv i   \bi{U}^\dagger(\vec x)  \left[  \bi{A}_i^{(\mathrm{q})}(t)\ ,
\bi{A}_n^{(\mathrm{q})}(t) \right] \bi{U}(\vec x)
.
$$
Enclosing the system in a proper quantization volume $V_H$
while keeping the time integration free
$$
\int_{V_H} d^3x\quad\longrightarrow \quad V_H
$$
we get
$$
S=V_H
\int dt \left[ {g_0{}^2\over 2N}  \mathrm{Tr} \left(  \bi{E}^i(t) \right)^2
-{1\over 2 }    \mathrm{Tr} \left( \bi{E}^i    {d\bi{A}_i\over dt} \right)-
{N\over 4 g_0{}^2}  \mathrm{Tr}  \left(   \left[ \bi{A}^{(\mathrm{q})}_i,
 \bi{A}^{(\mathrm{q})}_n \right] \right)^2  \right]
\ ,
$$
which is the action in the first order formulation; by substituting
for $\bi{E} ^{i} (t)$ its expression in terms of the vector potential
we obtain, in the second order formulation, an action
quite similar to the one for the bosonic sector of the BFSS model
describing a system of $N$ $D0$-branes in the gauge $A_0=0$:
\begin{equation}
S^{\mathrm{BFSS}}=V_H  {N\over g_0{}^2}
\int dt \left[ {1\over 2}
  \mathrm{Tr} \left( {d\bi{A}^{(\mathrm{q})}_i\over dt} \right)^2
-{1\over 4 }  \mathrm{Tr}  \left(   \left[ \bi{A}^{(\mathrm{q})}_i,
 \bi{A}^{(\mathrm{q})}_n \right] \right)^2  \right]
\ ;
 \label{abfss}
\end{equation}
the only
difference is the range of the spatial indices: we are working in three
rather than nine spatial dimensions. An action of the form (\ref{abfss}) can
also be obtained from monopole condensation and toroidal compactification
\cite{gabak}.

\section{Non-commutative Phase Space}
\label{moyal}
To match the large-$N$ $SU( N )$  gauge theory with some appropriate
brane model we have to bridge the gap between non-commuting  Yang--Mills
matrices and commuting brane coordinates. Since the world--trajectory of a
$p$-brane  is the target spacetime  image
$x^\mu=X^\mu(  \sigma^0,\sigma^1,\dots , \sigma^p )$ of the world
manifold $\Sigma : \sigma^m=(  \sigma^0,\sigma^1,\dots \sigma^p )$,
$X^\mu$ belonging to the algebra $\mathcal{A}$ of $C^\infty$ functions over
$\Sigma$, to realize our program we must \textit{deform} ${\mathcal{A}}$ to a
\textit{non-commutative ``starred'' algebra} by introducing a $\ast$-product.
The general rule is to define  the new product between two
functions  as  (for a recent review see \cite{harvey}):
\begin{equation}
f \ast  g =f \,g + \hbar\,  P_\hbar(\,  f\ ,g \,)
\ ,
\end{equation}
where $P_\hbar(\,  f\ ,g\, )$ is a bilinear map
$P_\hbar : {\mathcal{A}}\times {\mathcal{A}}\rightarrow
{\mathcal{A}}$. $\hbar$ is the \textit{deformation parameter} which is often
denoted by the same symbol as the Planck constant to stress the analogy
with quantum mechanics, where classically commuting dynamical variables
are replaced by non-commuting operators. In our case the role of deformation
parameter is played by
\begin{equation}
\hbar\equiv {2\pi\over N}
\ .
\end{equation}
For our purposes, we select $\Sigma={\mathbb{R}}^{2n}$ and choose the Moyal
product as the deformed $\ast$-product
\begin{equation}
    f(\sigma)  \ast   g(\sigma)
    \equiv
    \exp\left[  i  {\hbar\over 2} \omega^{mn}
    {\partial^2\over \partial  \sigma^m   \partial \xi^n}
    \right] f(\sigma)  g(\xi)
    \Biggr\rceil_{\xi=\sigma}
  \ ,
    \label{mp}
\end{equation}
where $\omega^{mn}$ is a non-degenerate, antisymmetric matrix, which
can be locally written as
\begin{equation}
\omega^{mn}=\left( \begin{array}{cc}
                 {\mathbb{O}}_{  n\times n}  & {\mathbb{I}}_{  n\times n}\\
                -{\mathbb{I}}_{  n\times n} & {\mathbb{O}}_{  n\times n}
                \end{array}
                  \right)
               \ .
\end{equation}
The Moyal product (\ref{mp}) takes a simple looking form in Fourier space
\begin{equation}
F(\sigma)  \ast   G(\sigma)=\int {d^{  2n}\xi\over (2\pi)^{n}}
\exp\left(  i  {\hbar\over 2} \omega_{mn}
\sigma^m    \xi^n \right)
F\left(  {\sigma \over 2}+ \xi  \right)
G\left(  {\sigma  \over 2}- \xi  \right)
\ ,
\end{equation}
where $F$ and $G$ are the Fourier transform of $f$ and $g$.
Let us consider the Heisenberg algebra
\begin{equation}
\left[  \bi{K}^m, \bi{P}^l \right]=i \hbar  \delta^{ml}\ ;
\end{equation}
Weyl suggested, many years ago, how an operator  ${\bi{O}}_F( \bi{K},
\bi{P} )$ can be written as a sum of  algebra elements as
\begin{equation}
\bi{O}_F= {1\over (2\pi)^n}\int d^np  d^nk  F(  p,k )
\exp\left(  i  p_m  \bi{K}^m + i   k_l  \bi{P}^l \right)
\ .
\label{weyltrasf}
\end{equation}
The Weyl map (\ref{weyltrasf}) can be inverted to associate functions, or
more exactly \textit{symbols}, to operators
\begin{equation}
F(\,  q\ ,k \,)= \int {d^n\xi\over (2\pi)^n} \exp\left(-i k \xi \right)
\Biggl\langle   q + \hbar{\xi \over 2}    \Biggl\vert  \bi{O}_F\left(\,\bi{K}\ 
,
 \bi{P} \,\right)  \Biggr\vert   q - \hbar {\xi \over 2 }  \Biggr\rangle
\ ;
\label{invweyl}
\end{equation}
moreover it translates the commutator between two
operators ${\mbox{\boldmath{$U$}}}$, ${\mbox{\boldmath{$V$}}}$ into the
\textit{Moyal Bracket} between their
corresponding symbols  ${\mathcal{U}}(\sigma)$, ${\mathcal{V}}(\sigma)$
\[
    {1\over i \hbar}   \left[   {\mbox{\boldmath{$U$}}},
    {\mbox{\boldmath{$V$}}}  \right]\longleftrightarrow
    \left\{{\mathcal{U}}, {\mathcal{V}}\right\}_{\mathrm{MB}}
    \equiv
    {1\over i  \hbar}
    \left(
    {\mathcal{U}} \ast {\mathcal{V}} - {\mathcal{V}} \ast {\mathcal{U}}
      \right)
\]
and the quantum mechanical trace into an integral over Fourier space
\begin{equation}
(2\pi)^n
    \mathrm{\mathrm{Tr}}_{{\mathcal{H}}} \bi{O}_F( \, \bi{K},
 \bi{P}\, ) \longmapsto
    \int d^{  n}p \,  d^{  n}k\, F( \, q\ ,k\,)  \equiv \int d^{2n}\sigma\,
   F( \, \sigma\,) \ .
\end{equation}
A concise but pedagogical introduction to
the deformed differential calculus and its application to
the theory of integrable system can be found in \cite{strach}.

We are now ready to formulate the alleged relationship between the quenched
model (\ref{abfss}) and membrane model: the symbol of the matrix
$\bi{A}^{(\mathrm{q})}_j$ is proportional to the $2n$-brane coordinate
$X^j(  \sigma^1 , \dots , \sigma^{2n} )$. Going through the steps discussed
above the action $S^{\mathrm{BFSS}}$ transforms into its symbol
$W^{\mathrm{BFSS}}$:
\begin{equation}
\fl S^{\mathrm{BFSS}}\rightarrow W^{\mathrm{BFSS}}=
{N  V_H\over 2\pi   g_0{}^2 }
\int dt\int d^{2n}\sigma\left[ {1\over 2 }
{d  A_i\over dt}  \ast   {d A^i\over dt}
+\frac{\left\{  A_i, A_n \right\}_{\mathrm{MB}}\ast
\left\{  A^i, A^n \right\}_{\mathrm{MB}}}{4}
\right]
\ .
\label{wbfss}
\end{equation}
The action (\ref{wbfss}) is manifestly Lorentz non-covariant, as it is
expected (the covariant, supersymmetric, higher dimensional version of the
action (\ref{wbfss}) is discussed in \cite{garcia}).
The adopted quenching scheme explicitly breaks the equivalence
between spacelike and timelike coordinates. Accordingly, our final result
takes a typical ``non-relativistic'' look.\\
Up to now we have not fixed the Fourier space dimension $n$. To
give $ A^i$   the meaning of embedding function, we have to choose
$2n\le D-1$, where $D$ is the target spacetime dimension. To match
QCD in four spacetime dimensions we set $n=1$. In this case
\begin{equation}
\omega^{mn}=\epsilon^{mn}=\left( \begin{array}{cc}
                               0 & 1\\
                               -1 & 0
                               \end{array}
                                \right)
\end{equation}
and we rescale the Yang--Mills charge and field\footnote{For the sake of
clarity, let us summarize the canonical dimensions in natural units of
various quantities:
\begin{eqnarray}
\fl [  A_\mu{}^a(x) ] \equiv
[ \bi{A}_\mu^{(\mathrm{q})} ]=(  \mathrm{length} )^{-1}\nonumber\\
\fl [  F_{\mu\nu}{}^a(x) ] \equiv
[ \bi{F}_{\mu\nu}{}^{(\mathrm{q})} ]=(  \mathrm{length} )^{-2}\nonumber\\
\fl [  \sigma^m ]=(  \mathrm{length} )^0=1\quad,\quad[
t ]=  \mathrm{length}\quad, \quad [ \beta ]=  \mathrm{length}  \nonumber\\
\fl [  g_0 ]\equiv [  g_{\mathrm{YM}} ]=(  \mathrm{length} )^0=1
\nonumber\\
\fl [  V_H ]= (  \mathrm{length} )^3  \nonumber\\
\fl [  X^i   ]=   \mathrm{length}\nonumber\\
\fl [  \mu_0 ]= (  \mathrm{length} )^{-1}\quad,\quad
[  \alpha ]= (  \mathrm{length} )^{-5}
\ .
\nonumber
\end{eqnarray}
%end of the footnote
}
as
\begin{eqnarray}
{N \over g_0{}^2} & \quad \longmapsto \quad {1 \over g^2_{\mathrm{YM}}}
\\
 A^i & \quad \longmapsto \quad V^{-2/3}_H\,  X^i
\ .
\end{eqnarray}
Since the ``glue'' is supposed to be confined inside an hadronic
size volume $V_H$, we can assign to $g_{\mathrm{YM}}$ the standard
value at the confinement scale
\begin{equation}
{g^2_{\mathrm{YM}}\over 4\pi} \simeq 0.18
\ .
\end{equation}
Finally, if $N \gg 1$ the Moyal bracket can be approximated by the Poisson
bracket
$$
\left\{  X^i, X^j \right\}_{\mathrm{MB}}
\longmapsto
\left\{  X^i, X^j \right\}_{\mathrm{PB}}
$$
and (\ref{wbfss}) takes the form \cite{neman}
\begin{equation}
S^{p=2}_{\mathrm{NR}}=
\int dt  d^2\sigma  \left[ {1\over 2 }  \mu_0
  {d X_k\over dt}    {d X^k\over dt}
+{\alpha\over 4 }  \left\{  X_i, X_j \right\}_{\mathrm{PB}}
\left\{  X^i, X^j \right\}_{\mathrm{PB}}  \right]
\ ,
\label{mnr}
\end{equation}
where $\mu _{0}$ and $\alpha$ are defined by
$$
\mu_0\equiv { V_H^{-1/3}\over 2\pi  g^2_{\mathrm{YM}}  }
\quad , \quad
\alpha\equiv { V_H^{-5/3}\over 2\pi  g^2_{\mathrm{YM}}  }
\ .
$$
Moreover from the definition of the Poisson bracket
$$
\left\{  X^i, X^j \right\}_{\mathrm{PB}}
\equiv
\epsilon^{mn}\partial_m  X^i  \partial_n  X^j
$$
we can compute
$$
\left\{  X^i\ , X^j \right\}_{\mathrm{PB}}\left\{  X_i\ , X_j
\right\}_{\mathrm{PB}
}=2\cdot\det\left(\,  \partial_m \,  X^k \, \partial_n \,  X_k \,\right)\equiv 
2
\gamma
$$
so that the action (\ref{mnr}) can be rewritten as
\begin{equation}
S^{p=2}_{\mathrm{NR}}=
\int dt  d^2\sigma  \left[ {1\over 2 }  \mu_0
  {d X_k\over dt} \,   {d X^k\over dt}
+{\alpha\over 2 }  \det \left( \, \partial_m \,  X^k \, \partial_n \,  X_k
\,\right)\,
 \right]
\ .
\label{nrm}
\end{equation}
The first term in (\ref{nrm}) is a straightforward generalization of the
kinetic energy of a non-relativistic particle; the second term represents
the ``potential energy'' associated  to the elastic deformations of the
membrane. The action (\ref{nrm}) still displays a residual symmetry under
\textit{area preserving diffeomorphisms}, leaving  only one dynamical degree of
freedom describing transverse oscillations of the membrane
surface\footnote{We have assumed that $X^i$ are three spacelike coordinates. By
relaxing this assumptions we can give  (\ref{nrm}) a slightly different
physical interpretation. If  all three $X^i$ are considered as
\textit{transverse} directions,
then (\ref{nrm}) can be seen as the \textit{light--cone gauge}
action for a bosonic brane in a $5$-dimensional target spacetime.}.
If the action (\ref{nrm}) has any chance to provide a membrane model of
hadronic objects, then it must be able to provide at least the correct
order of magnitude of hadronic masses. Our model does not take into account
spin effects, therefore it is consistent to look for spherically symmetric
configurations. Again this is a sort of quenching even if of a more
geometric type. Infinite vibration modes of the brane, corresponding to local
shape deformations, are frozen and the dynamics is reduced to the ``radial''
breathing mode alone.
This kind of approximation, commonly called ``minisuperspace'' approximation,
is currently adopted in Quantum Cosmology, where
it amounts, in practice, to quantize a single scale factor (thereby
selecting a class of cosmological models, for instance, the
Friedman--Robertson--Walker
spacetimes) while neglecting the quantum fluctuations of the full
metric. The effect is to turn the exact, but intractable,
Wheeler--DeWitt
functional equation \cite{wdw} into an ordinary quantum
mechanical wave equation \cite{padma}.
As a matter of fact, the various forms of the ``wave function of the
universe'' that attempt to describe the quantum birth of the cosmos are
obtained through this kind of   approximation \cite{hh}
or modern refinements of it \cite{gianni}. This non-standard
approximation scheme was  applied to a
relativistic membrane  in the seminal paper by Collins and Tucker
\cite{coll}, and since then it has been used several times \cite{mini},
including the case of self--gravitating objects \cite{bolle}.

%%%%%%%%%%%%%%%%%%%%%%%%%%%%%%%%%%%%%%%%%%%%%%%%%%%%%%%%%%%%%%%%%%%%%%%%%%%%%%%
%

Following \cite{coll}, we parametrize the membrane coordinates as follows
\begin{eqnarray}
X^1\equiv R(t) \sin\theta\cos\phi\nonumber\\
X^2\equiv R(t) \sin\theta\sin\phi\label{emb}\\
X^3\equiv R(t)\cos\theta\nonumber
%.
\end{eqnarray}
and the transverse, dynamical degree of freedom corresponds to $R$.
The metric $\gamma_{ab}$ induced on the membrane by the embedding (\ref{emb})
is:
$$
\gamma_{ab}= \mathrm{diag}\left(  R^2(t), R^2(t) \sin^2\theta 
\right)\quad,\quad
\det\left(  \gamma_{mn} \right)=  R^4(t) \sin^2\theta
\ .
$$
The corresponding action turns out to be
$$
S=
\int L dt
=
\pi^2\int dt  \left[  {1\over 2}  \mu_0  \left( {dR\over dt} \right)^2 +
{\alpha\over 4}  R^4 \right]
\ ;
$$
accordingly the momentum conjugated to the only dynamical degree of
freedom is
$$
P_R\equiv {\partial L\over \partial (  dR/dt  )}=\pi^2   \mu_0
{dR\over dt}
$$
from which the Hamiltonian can be calculated as
$$
H\equiv P_R   {dR\over dt} - L ={1\over 2\pi^2 \mu_0}  P_R^2  +
{\alpha\pi^2\over 4}  R^4
\ .
$$
Then the action in Hamiltonian form is
$$
S=\int dt  \left[  P_R  {dR\over dt}   - \left(  {1\over 2\pi^2 \mu_0}P_R^2
+ {\alpha\over 4}  R^4 \right) \right]
\ .
$$
The above results allow one to compute the hadronic
mass spectrum from the spherical membrane Schr\"odinger equation
\begin{equation}
\left[  -{1\over 2\pi^2 \mu_0}{d^2\over dR^2} + {\alpha\pi^2\over 4}  R^4
 \right]  \Psi(  R )= M_n \, \Psi(  R )
\end{equation}
with the following boundary conditions:
\begin{eqnarray}
\Psi(  0 )=0\\
\lim_{R\to\infty}\Psi(  R )=0
.
\end{eqnarray}
The lowest mass eigenvalues can be evaluated numerically \cite{eigen} or
through WKB approximation \cite{mini}:
\begin{equation}
\left( 2\pi^2 \mu_0  M_n \right)^{1/2}={4\pi\over\beta(1/4, 3/2)}
 \left( {1\over 8\sqrt 2 V_H  g_{\mathrm{YM}}^2 } \right)^{1/3}
 \left(  n +{3\over 4} \right)^{2/3}
\ .
\label{mwkb}
\end{equation}
where, $\beta$ is the Euler $\beta$--function: $\beta(1/4, 3/2)\equiv
{\Gamma(1/4)\, \Gamma(3/2)\over\Gamma(7/4)}$.
The WKB formula (\ref{mwkb})  gives a mass scale of the
correct order of magnitude, $ M_n \propto 4\pi\, g_{\mathrm{YM}}^{2/3}V_H^{-1/3
}\simeq 1 \mathrm{GeV}$. More sophisticated estimates of the glueball mass
spectrum, including topological corrections \cite{gaba}, are not very different
from the values given by (\ref{mwkb}). Thus, we conclude that the QCD
membrane action (\ref{nrm}) encodes, at least the dominant contribution,
to the gluon bound states spectrum.
%%%%%%%%%%%%%%%%%%%%%%%%%%%%%%%%%%%%%%%%%
Hopefully, an improvement of this result will come from an extension of the
minisuperspace approximation along the line discussed in \cite{prop}, where
a new form of the p--brane propagator has been obtained.

\end{document}